\documentclass[floatfix,%
 aip,
 amsmath,amssymb,
 preprint,%
 reprint,%
]{revtex4-1}

\usepackage{graphicx}
\usepackage{dcolumn}
\usepackage{bm}
\usepackage{xcolor}
\usepackage{booktabs}

\usepackage[utf8]{inputenc}
\usepackage[T1]{fontenc}
\usepackage{mathptmx}

\begin{document}

\title{Interpreting AI for Fusion: An Application to Plasma Profile Analysis for Tearing Mode Stability}

\author{Hiro J Farre-Kaga}
\affiliation{The two authors contributed equally to this paper}
\affiliation{Princeton University, Princeton, NJ, USA}
\affiliation{Princeton Plasma Physics Laboratory Princeton, NJ, USA}
\author{Andrew Rothstein}
\affiliation{The two authors contributed equally to this paper}
\affiliation{Princeton University, Princeton, NJ, USA}
\author{Rohit Sonker}
\affiliation{Carnegie Mellon University, Pittsburgh, PA, USA}
\author{SangKyeun Kim}
\affiliation{Princeton Plasma Physics Laboratory Princeton, NJ, USA}
\author{Ricardo Shousha}
\affiliation{Princeton Plasma Physics Laboratory Princeton, NJ, USA}
\author{Minseok Kim}
\affiliation{Princeton University, Princeton, NJ, USA}
\author{Keith Erickson}
\affiliation{Princeton Plasma Physics Laboratory Princeton, NJ, USA}
\author{Jeff Schneider}
\affiliation{Carnegie Mellon University, Pittsburgh, PA, USA}
\author{Egemen Kolemen}
\affiliation{Princeton University, Princeton, NJ, USA}
\affiliation{Princeton Plasma Physics Laboratory Princeton, NJ, USA}

\begin{abstract}
AI models have demonstrated strong predictive capabilities for various tokamak instabilities—including tearing modes (TM), ELMs, and disruptive events—but their opaque nature raises concerns about safety and trustworthiness when applied to fusion power plants. Here, we present a physics-based interpretation framework using a TM prediction model as a first demonstration that is validated through a dedicated DIII-D TM avoidance experiment. By applying Shapley analysis, we identify how profiles such as rotation, temperature, and density contribute to the model's prediction of TM stability. Our analysis shows that in our experimental scenario, a large density profile is lightly destabilizing, but core electron temperature and rotation peaking play the primary role in TM stability. This work offers a generalizable ML-based event prediction methodology, from training to physics-driven interpretability, bridging the gap between physics understanding and opaque ML models.
\end{abstract}
\keywords{Machine Learning, Interpretable AI, Tearing Modes, Tokamak, DIII-D, Plasma Control}

\maketitle

\section{Introduction}\label{sec:intro}
Tokamaks are a promising fusion energy technology, but they face challenges in maintaining plasma stability. Recently, machine learning (ML) and artificial intelligence (AI) have been applied more and more to the field of nuclear fusion in the form of surrogate physics models \cite{morosohkAcceleratedVersionNUBEAM2021,boyerRealtimeCapableModeling2019b,morosohkNeuralNetworkModel2021,rothsteinInitialTestingAlfven2024}, event prediction models \cite{keithRiskAwareFrameworkDevelopment2024a,seoMultimodalPredictionTearing2023b,aiDisruptionPredictorBased,reaRealtimeMachineLearningbased2019a,jalalvandAlfvenEigenmodeClassification2022,churchillDeepConvolutionalNeural2020a}, and reinforcement learning controllers \cite{seoAvoidingFusionPlasma2024b,degraveMagneticControlTokamak2022a}. However, a key requirement for next-generation fusion power plants is to have interpretable control systems where causes of control actions can be directly tied to observations in the plasma. This is directly at odds with the typical AI approach that utilizes the high prediction accuracy of black-box models that are uninterpretable. 

Disruption prediction approaches have utilized "gray-box" models, such as random forest models \cite{reaRealtimeMachineLearningbased2019a}, that offer interpretable results at the trade-off of simpler ML architectures with lower accuracies. However simple black-box models such as multilayer perceptrons (MLP) have the advantage of ease of training, allowing researchers to produce highly accurate machine learning models with fewer resources and expertise. Interpretable neural networks may require restricting the model's architecture and number of parameters, which can lead to lower accuracy than a deep complex network~\cite{liuInterpretableNeuralNetworks2023}

Instead of changing the "black-box" model architectures to gain interpretability, we can adjust our analysis framework to gain insights from these "black-box" models using Shapley analysis. Shapley analysis is a method for explaining the output of machine learning models based on a game theoretic approach \cite{lundbergUnifiedApproachInterpreting2017} by fairly distributing the prediction result across model inputs. This analysis framework can be applied to any model, machine learning-based or otherwise. 

In this application, we study how the plasma profiles affect tearing mode (TM) stability and explain which specific profile features, such as rotation peaking, can stabilize TMs. Previous TM prediction and stability models used just scalar parameters \cite{olofssonEventHazardFunction2018a,olofssonHazardFunctionExploration2019,fuMachineLearningControl2020c}, a subset of magnetic field features \cite{Olofsson_Akçay_Sammuli_2022}, or full plasma profiles with no stability interpretation \cite{seoMultimodalPredictionTearing2023b}. We improve on these with a ML-based Deep Survival Machine model \cite{nagpalDeepSurvivalMachines2021c} to predict TMs based on real-time plasma profiles. Using Shapley analysis, we can explore how the plasma profiles affect TM stability and explain what specific profile features, such higher core $T_e$ and $T_i$, led to the avoidance of TMs. This model is applied to a dedicated TM avoidance experiment on DIII-D, and its results are used for this analysis. While Shapley analysis has been applied to other fields\cite{lundbergExplainableMachinelearningPredictions2018}, there are minimal applications to fusion and has not been used for understanding fusion experiments \cite{landremanHowDoesIon2025a,pyragiusApplicationInterpretableMachine2024}, as far as the authors are aware.

TM stability analysis poses an interesting problem for interpretation as many physics studies have been performed to better understand these instabilities \cite{bardoczi_onset_2023,bardoczi_empirical_2023}, often with conflicting results or limited to scenario-specific operating regimes such as the DIII-D ITER baseline\cite{turcoCausesDisruptiveTearing2018,bardocziRootCauseDisruptive2024,richnerUseDifferentialPlasma2024a}. Other approaches to better understand TM stability involve using a physics code such as STRIDE to calculate the the classical $\Delta^\prime$ stability parameter \cite{glasserIdealMHDDW2020a}, however, this has not been validated in experiments. The results from analyzing our TM prediction model with Shapley analysis can add additional information to the greater plasma physics discussion of TM stability. 

This paper begins with an explanation of the tools and techniques used for training and interpreting the TM predictors in Section~\ref{sec:background}. Section~\ref{sec:performance} describes the TM predictor model results, followed by interpretation for the TM preemptive avoidance experiment. Then we use Shapley values to draw conclusions more broadly about profile-based TM stability. Finally Section~\ref{sec:conclusion} summarizes our findings and describes the future work to improve TM models and their interpretation. 

\begin{figure*}
    \centering
    \includegraphics[width=170mm]{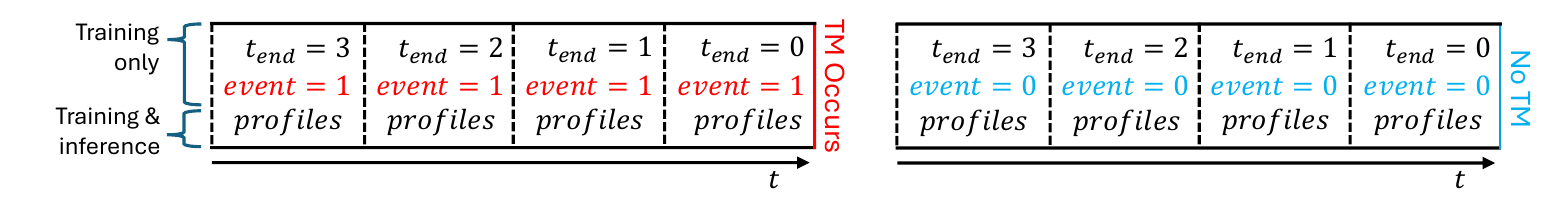}
    \caption{Depiction of the survival regression training scheme. In training, the model is input $t_{end}$ representing the time until the end of the sequence, $event$ representing whether a TM occurs at the end of the sequence (1) or not (0), and $profiles$ representing the set of diagnostics and inputs to the model. At inference, only $profiles$ are input, since the end of sequence or event are of course not known.}
    \label{fig:auton}
\end{figure*}

\section{Background}\label{sec:background}

We begin with a description of the database processing used to train our TM prediction model in Section~\ref{data_processing}, followed by an explanation of the training method in Section~\ref{auton} and the Shapley model interpretation in Section~\ref{shap_section}.

\subsection{Database processing and TM labelling}
\label{data_processing}

The model was trained on all DIII-D shots identified to have the required data between shots 140000 to 195000, resulting in 6050 shots of which 1476 contained n=1 TMs, where n is the toroidal mode number, which amounts to 677494 timesteps each with 42 parameters. The data needed is listed in Table \ref{tab:input-parameters} and includes Thomson Scattering~\cite{carlstrom_design_1992}, Charge Exchange Recombination Spectroscopy~\cite{seraydarian_multichordal_1986}, Motional Stark Effect~\cite{wroblewski_motional_1990}, magnetics for EFIT reconstructions~\cite{lao_reconstruction_1985} and actuation values such as neutral beam and electron cyclotron heating power. While training scenario-specific models may improve performance in that scenario, the aim of this model is to be flexible so it may be applied to different scenarios for DIII-D experiments. 

Importantly, our dataset has no differentiation between classical TMs and neo-classical TMs (NTMs) since they both appear similarly in our automated labeling. However, the planned control actions should be effective for both TM and NTM stabilization by replacing the missing bootstrap current. Consequently, when we refer to TMs it is assumed to include both classical and neo-classical TMs. We also do not consider locked or quasi-stationary modes and limit our analysis to born-rotating modes, as locked modes are often preceeded by born-rotating modes~\cite{sweeney_relationship_2018}. 

The following are the key database processing steps taken and their rationale, with further details covered in Appendix~\ref{app:data_processing}:

\begin{itemize}
    \item Plasma current ($I_P$) rampup and rampdown are excluded as we are only targeting TMs in $I_P$ flattop. 
    \item A TM was considered to have occurred if the amplitude of the root mean square (RMS) of the n=1 magnetic fluctuations peaked above 12G for a continuous 50ms along with additional constraints on $H_{98}$ and $q_{95}$ to only consider H-mode plasmas. The onset time of the TM was determined to be when the n=1 RMS amplitude first reached 10\% of the peak n=1 RMS amplitude.
    \item Magnetic fluctuation signals like the n=1,2,3 magnetic RMS signals that identify the n=1,2,3 modes are excluded from model inputs in the training set to avoid the model overly relying on these signals as they are used for labeling.
    \item The data is taken every 20ms as this is enough time for diagnostics and EFITs to yield updated results, faster than $\tau_R$ and $\tau_E$ ensuring the profile are equilibrated, but not too fast that the model overfits to noise. Actuation such as changes to ECH deposition and NBI power adjustment will also affect the plasma on the order of $100ms$, so this is a good compromise.
\end{itemize}

\subsection{The survival regression training scheme}
\label{auton}

The model in this paper uses the Deep Survival Machines (DSM) architecture from the open-source Auton-Survival package \cite{nagpalDeepSurvivalMachines2021c}. The framework allows for easy-to-use event prediction, and has been proven in fusion applications for disruption prediction\cite{keithRiskAwareFrameworkDevelopment2024a} to achieve longer warning times compared to other models. Like any event prediction model, the two key ingredients are accurate labels and input data that is representative of the underlying physics. Fig~\ref{fig:auton} depicts the training scheme, where in training we input the plasma parameters such as the profiles, the time until the end of sequence, and whether or not a TM occurs at the end of the sequence. At inference, or when running the model, we only input the plasma parameters since the event or time-to-event are not known. This training setup informs the model whether the given plasma parameters will be unstable $t$ timesteps in the future.

Survival regression is a statistical scheme that provides a probability of an event occurring at any time within a user-chosen time horizon, $t_{horizon}$, given a set of input features. A common application of this algorithm is in estimating the survival times of patients given certain treatments and symptoms, hence the name survival. By analogy, a TM in a plasma may be considered a 'death', and the input features such as the density and temperature as the 'symptoms'. This is therefore applicable to plasma disruptions in general, and to specific MHD events such as the onset of a $m,n=2/1$ TM for poloidal mode number $m$ and toroidal mode number $n$. 

\subsection{Shapley analysis for model interpretation}\label{shap_section}

Machine learning models such as the above survival regression are often considered black boxes, as they consist of matrix multiplication sequences with millions of uninterpretable parameters. While these matrix weights are difficult to justify, we can still understand and explain a black-box model by studying how the input features affect the output. For example, a sudden drop in rotation may lead to a TM, so such an input change should increase the model's TM probability. Similarly, this analysis provides an insight into \textit{why} the model predicts a TM, and what plasma feature was most responsible. 

Shapley analysis is a game theoretic approach~\cite{lundbergUnifiedApproachInterpreting2017} to analyze the impact of input parameters on the model output.  Shapley values represent the contribution of a particular feature value to the overall prediction of the model relative to a background distribution. The Shapley value corresponding to an input represents its contribution to the model output relative to the average effect in the background distribution. Hence, the total Shapley value equals the model's output minus the model's average output in the background distribution. To demonstrate this technique, we have provided an illustrative toy example in Appendix \ref{app:toy-model}. 

In our TM study, we use the 11 shots from our dedicated DIII-D experiment as the background distribution. Another choice for the background would be the entire DIII-D dataset, but narrowing this down to the experiment shots allows for a more in-depth comparison between the 11 shots. For example, $\beta_N$ will not be a significant factor in shots from our experimental session that achieved similar $\beta_N\sim 3$, but if we were to compare to standard DIII-D H-mode shots with $\beta_N\sim 2$, the effect of $\beta_N$ would be overwhelming. Thus in our evaluation of TM predictions later, it is important to ground these interpretations based on the relevant reference distribution of plasma equilibria. 

\begin{table}[tbp]
    \centering
    \begin{tabular}{@{}ll@{}}
    \toprule
    \textbf{Input} & \textbf{Source} \\
    \hline
    \midrule
    Electron Temperature profile ($T_e$)       & RTCAKENN \\
    Electron density profile ($n_e$)           & RTCAKENN \\
    Ion temperature profile ($T_i$)            & RTCAKENN \\
    Ion rotation profile ($Rot$)         & RTCAKENN \\
    Plasma Pressure profile ($p$)                    & RTCAKENN \\
    Safety factor profile ($q$)               & RTCAKENN \\
    Current density profile ($j$)             & RTCAKENN \\
    NBI power                               & Neutral Beam Injection \\
    NBI torque                              & Neutral Beam Injection \\
    ECH power                               & Electron Cyclotron Heating \\
    $I_P$                                      & Plasma Current \\
    $B_T$                                   & Toroidal magnetic field \\
    Normalized pressure ($\beta_n$)         & EFITRT2  \\
    $q_{\min}$                              & EFITRT2 \\
    Internal inductance ($l_i$)            & EFITRT2 \\
    Plasma minor radius ($a_{\text{minor}}$)& EFITRT2 \\
    Plasma major radius ($R$)              & EFITRT2 \\
    Bottom Triangularity ($\delta_{\text{bot}}$) & EFITRT2 \\
    Top Triangularity   ($\delta_{\text{top}}$) & EFITRT2 \\
    Elongation ($\kappa$)                  & EFITRT2 \\
    Plasma volume (Vol)                    & EFITRT2 \\
    \bottomrule
    \end{tabular}
    \caption{Input parameters used in the analysis and their corresponding sources. EFITRT2 is real-time magnetic equilibrium reconstruction using magnetic diagnostics and motional stark effect. All profiles from RTCAKENN are reduced to the 4 main PCA components. }
    \label{tab:input-parameters}
\end{table}

\begin{figure}
    \centering
    \includegraphics[width=74mm]{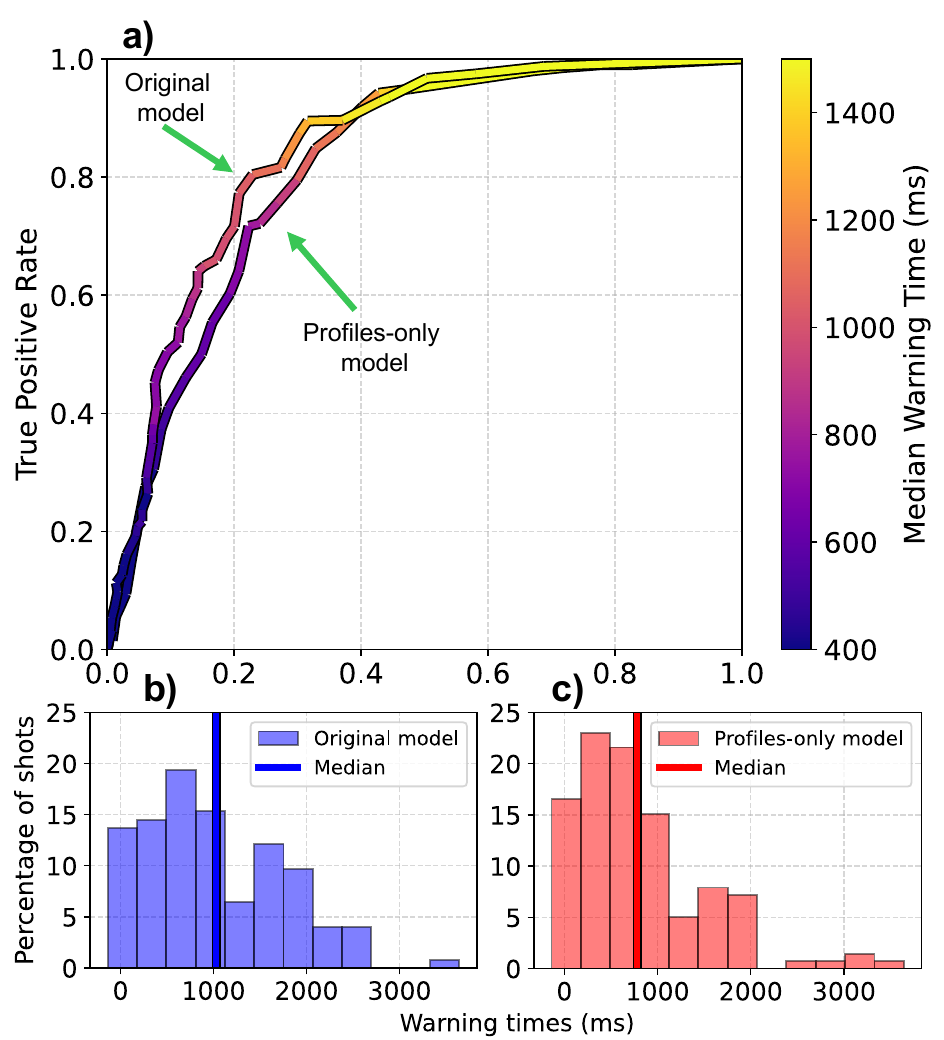}
    \caption{a) ROC curve for the RTCAKENN-based TM predictor used in the experiment. The original model had an AUROC score of 0.85, compared to 0.82 for the profiles-only model. b) Warning times histogram of true positives for a typical threshold of 0.2 shows the majority of TMs are predicted over 500ms in advance, allowing for flexibility in actuation.}
    \label{fig:roc_curve}
\end{figure}

\begin{figure*}[t]
    \centering
    \includegraphics[width=\textwidth]{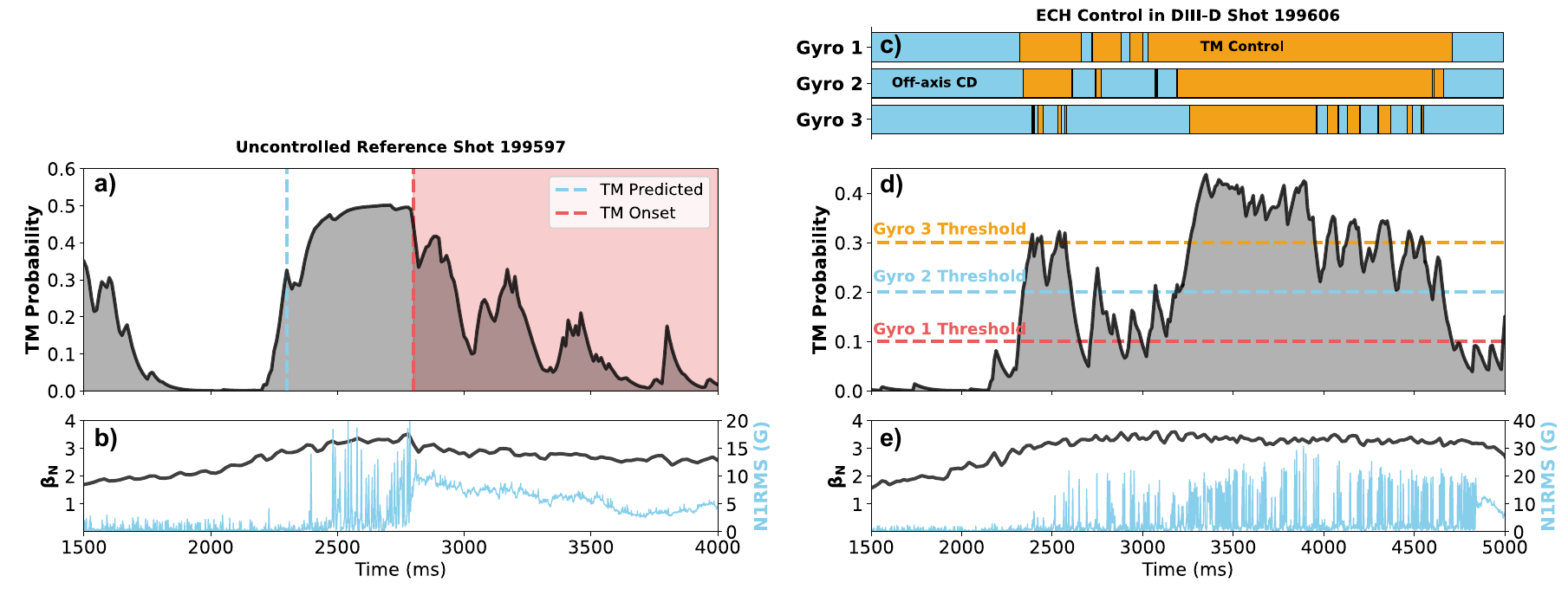}
    \caption{Demonstration of active preemptive TM suppression via ECH deposition change. a),b) shows the uncontrolled reference shot, which resulted in a TM at t=2700ms, predicted at t=2300ms. c), d), e) shows a shot with TM suppression via ECH deposition change. In c), blue represents ECH aimed at an off-axis location and orange represents ECH aimed at the q=2 surface for TM control.}
    \label{fig:experiment}
\end{figure*}

\section{Model performance and physics interpretation}\label{sec:performance}

This section describes the application of the TM prediction model, from its performance statistics in the DIII-D TM database to the model's application to a dedicated control experiment, and finally an analysis of the plasma profile features impacting TM stability. We begin listing the input features to the model and presenting its performance metrics in Section~\ref{tm_model}. The DIII-D experiment on preemptive TM avoidance is explained in detail in Section~\ref{experiment}, showing successful TM avoidance and detailing the model's results shot-by-shot. We look into specific time-slice profiles and discuss the interpretation of the model using Shapley analysis in Section~\ref{sec:shap1}, where the actuator effects on equilibrium profiles are shown to stabilize TMs. Finally in Section \ref{sec:shap2} we study the Shapley values across our experiment to draw broader conclusions on the scenario's stability.

\subsection{Tearing mode prediction model}\label{tm_model}
A TM prediction model was trained to predict DIII-D n=1 TMs using the parameters shown in Table~\ref{tab:input-parameters}, including the real-time kinetic profiles RTCAKENN~\cite{shoushaMachineLearningbasedRealtime2023}, an ML surrogate model for CAKE~\cite{xingCAKEConsistentAutomatic2021}, as well as external heating and actuation, and EFITRT2 scalars. The decisions on database selection and processing are listed in \ref{data_processing}. Most decisions were made to fit the experimental design, which required a real-time capable flattop TM predictor with enough warning time to change the electron cyclotron heating (ECH) deposition and affect the equilibrium (around 200ms).

The basic performance metrics are shown in Fig.~\ref{fig:roc_curve}, demonstrating high performance metrics with warning times around 1000ms an AUROC score of 0.85. This allows for flexibility in actuation to avoid the oncoming TM, such as ECH deposition in the case of our experiment. Further notes on the definition of warning times and classification details are given in Appendix \ref{app:event-char}.

The model architecture was a Deep Survival Machine with the following parameters: a 80/10/10 train/validation/test split, an MLP with four layers of dimension $128\times128$, a log-normal distribution, batch size of 256, learning rate of $2\times10^{-4}$, $k=2$ survival distributions, dropout of 0.4 and 100 epochs. The train/validation/test split was done on a per-shot basis to avoid inflating performance metrics where potentially similar timeslices from a single shot could end up in all three groups. 

A second model whose inputs are only the 7 RTCAKENN profiles was also trained but not used in experiment. This model is used in Section~\ref{sec:shap1} to study how profile changes affect the TM risk without confounding scalar inputs. Specifically, $\beta_N$, input heating and some shape parameters were found to have consistent, large Shapley values which made analysis of profile importance more difficult. This model of course has worse performance metrics as it used fewer diagnostics and inputs, but the AUROC score of 0.82 is not significantly worse for the purpose of the Shapley analysis. A difference in warning time distribution between the two models can be seen in Fig.~\ref{fig:roc_curve}b),c). 

\subsection{DIII-D experiment: preemptive tearing mode suppression}\label{experiment}

The TM predictor was developed for a dedicated DIII-D experiment to achieve active 2/1 TM suppression, consisting of aiming electron cyclotron current drive (ECCD) at the $q=2$ surface when a TM risk was predicted. Whenever ECH is applied in this experiment, our experimental setup provided ECCD as well. We therefore refer to the combined effect as ECH and differentiate these when referring to just the heating or current drive effect. The experiment was run in the elevated $q_{min}$ scenario~\cite{victorGLOBALSTABILITYELEVATEDQMINb}, a high-performance advanced non-inductive scenario that is often limited by TMs. This publication \cite{rothsteinaHighqmin} by the authors of this paper describes the experiment and its results in detail.

Previous experiments have shown the potential of preemptive suppression of TMs using ECCD\cite{kolemenStateoftheartNeoclassicalTearing2014d} by changing the ECCD deposition when TMs are detected. However 2/1 modes are difficult to suppress once they have appeared, and cause significant performance degradation while they are present. We therefore designed a preemptive scheme, where we predict TMs before they appear, and change the ECH deposition to stabilize the profiles, enabling fully tearing free operation. The experiment successfully demonstrated the suppression scheme as can be seen in Fig.~\ref{fig:experiment}, resulting in the tearing free operation of previously unstable conditions. 

\begin{figure*}
    \centering
    \includegraphics[width=0.9\textwidth]{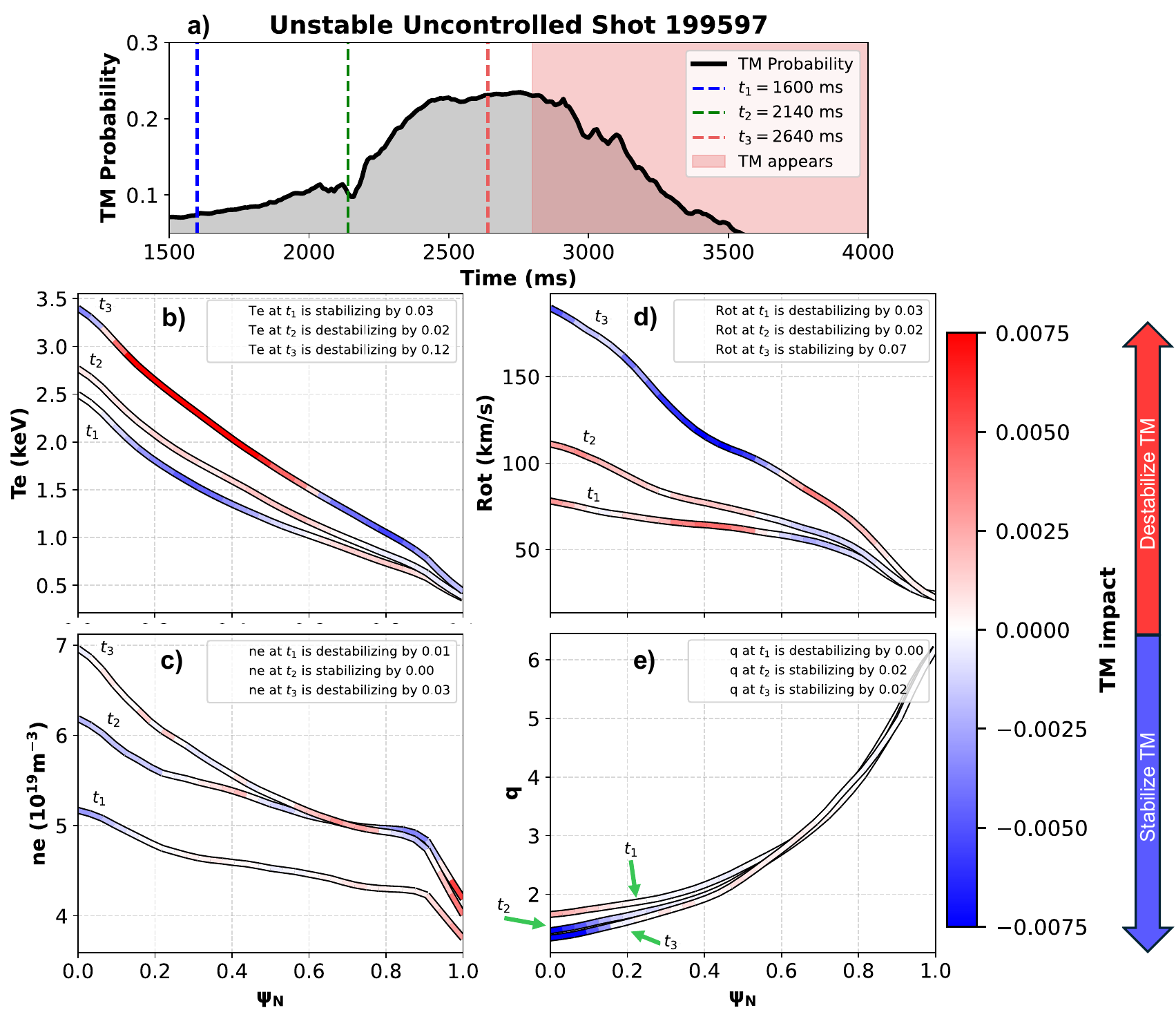}
    \caption{Shapley analysis for the unstable uncontrolled reference shot 199597. 3 timesteps are chosen, $t_1 = 1600$ ms at the stable start of the shot, $t_2 = 2140$ ms right before the TM is predicted, and $t_3 = 2640$ ms where tearing risk peaks and a TM occurs at $t = 2700$ ms. The color of the profiles represents the tearing impact, where red is destabilizing (positive TM impact), blue is stabilizing (negative TM impact), and white has no impact. The total impact on stability of the profile, or the sum of Shapley values, is in the legend. Abbreviations are used from Table \ref{tab:input-parameters}. }
    \label{fig:shap_199597}
\end{figure*}

\begin{figure*}
    \centering
    \includegraphics[width=0.9\textwidth]{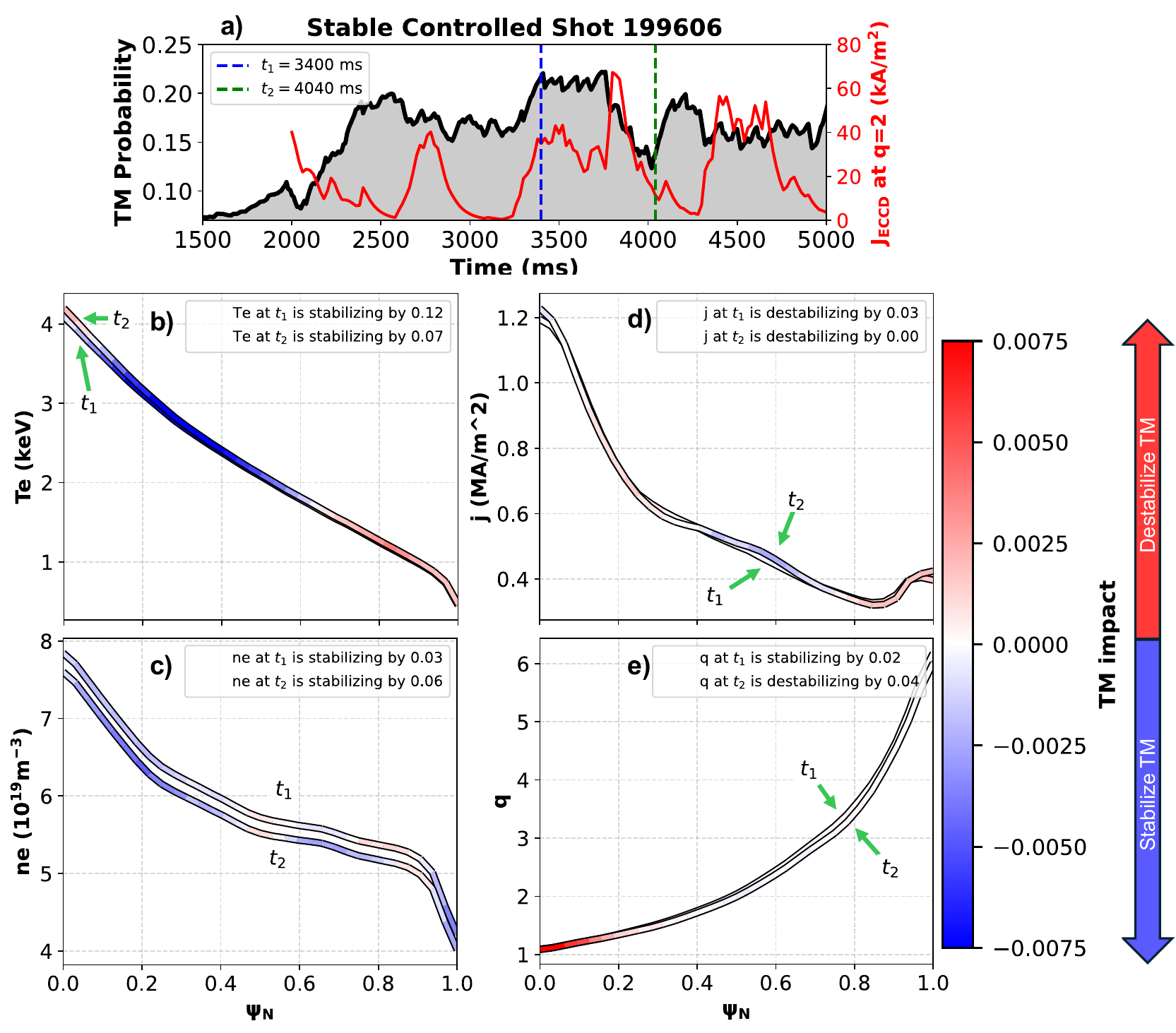}
    \caption{Shapley analysis for the ECH controlled stable shot 199606. Two timesteps are chosen, $t_1 = 3140ms$ when $J_{ECCD}$ at $q=2$ is low and the tearing risk is high, and $t_2 = 3920ms$ where ECH deposition has been changed so $J_{ECCD}$ at $q=2$ is high and the tearing risk begins to drop. The rotation profile is not shown, as the small difference between $t_1$ and $t_2$ had little impact.}
    \label{fig:shap_199606}
\end{figure*}

More importantly for this paper, the TM model successfully predicted the TMs with sufficient warning time to enable actuation, and responded correctly to the stabilizing effects of modifying ECH deposition. The experiment started at shot 199597 which was a reference elevated $q_{min}$ shot, and ended at 199607. The reference was well predicted, as shown in Fig.~\ref{fig:experiment}, followed by another correct unstable prediction in 199598 and 199599. Shots 199605, 199606, 199607 were the same unstable conditions with predicted TMs, but the active adjustments of ECH deposition successfully avoided TMs. Shots 199600 and 199601 used additional ECH power which led to passively stable conditions, correctly predicted again. The only failure of the predictive model was shots 199602 and 199603 which were run at lower plasma current, creating n=3 modes that triggered n=1 TMs. This resulted in a 82\% success rate, where the 2 failures were difficult for our model to predict as higher $n$ modes drove n=1 modes, something our current model is unable to account for, as larger n modes do not have significant profile flattening effects. Future predictor models could incorporate information from magnetic fluctuation signals to have information about higher n modes and their triggering of n=1 modes. 

\subsection{Shapley analysis I: what drove the TM, and how was it suppressed?}\label{sec:shap1}
Shapley analysis is performed on the two key shots of our experiment shown in Fig.~\ref{fig:experiment}: shot 199597 as the no-control baseline, where a TM occurred and was predicted 500ms in advance; and shot 199606, which had the same actuation and conditions as the reference, with the sole difference that ECH deposition is changed from an off-axis current drive position to the q=2 surface whenever TMs were predicted. 

For the following analysis, we use the new profile-only model whose inputs were solely the 7 profiles defined in table \ref{tab:input-parameters}, which explains the small differences in TM probability predictions for the same shots between Fig.~\ref{fig:experiment} and Figs. ~\ref{fig:shap_199597} and~\ref{fig:shap_199606}. We aim to understand how the equilibrium profiles determine the stability of the plasma; therefore the actuation scalars were not included in the model as they indirectly impact the stability by changing the profiles. While Shapley analysis may be performed on the original model, it is important to remove highly correlated values such as $\beta_N$, which is correlated to the pressure profile, to avoid ambiguity in Shapley values. Using only profiles allows a study of their true impact on TM stability. 

For this analysis, we refer to the calculated "Shapley value" as "TM impact" to make the interpretation of the values explicitly clear. In Figs. \ref{fig:shap_199597} and \ref{fig:shap_199606}, a positive TM impact (positive Shapley value) describes TM destabilization and is represented by redder colors, while negative TM impact (negative Shapley value) causes TM stabilization and is represented by bluer colors.  

The four profiles types shown in Fig.~\ref{fig:shap_199597} reflect the evolution of the plasma, and its impact on TM risk. The general increase in $T_e$ leads to more positive TM impact, particularly in the off-axis region ($0.3 < \psi_N < 0.8$). The pedestal growth in the edge region ($0.8 < \psi_N$) has a stabilizing effect seen with the negative TM impact values. This is generally consistent with the understanding that higher temperatures and pressures will lead to more tearing risk. 

Similarly, the evolution of TM impact values for the rotation profile provides an insight into the causes of the observed TM. As the core rotation increases, the region becomes more stabilizing as would be expected from an MHD stability perspective. The higher edge rotation has the opposite, slightly destabilizing effect, which suggests that rotation gradient or peaking is an important feature. The net change in the rotation profile's effect on TM risk (shown in the figure legend) from $t_1$ to $t_3$ is a stabilizing 0.1, which is lower than the the 0.15 destabilizing impact of the $T_e$ increase, suggesting that this TM is primarily caused by high core $T_e$. 

The density profile evolution has a relatively low impact to TM probability, but has interesting features. The electron density profile is elevated in $t_2$ and $t_3$ compared to $t_1$ even though $t_3$ is more destabilizing than $t_1$ or $t_2$. This suggests the peaked shape in $t_3$ is responsible for the destabilizing effect rather than elevated average densities. Finally, while the $q$ profile evolution had little influence on tearing, the figure is included as it shows interesting physics information. The drop in $q_{min}$ has a stabilizing effect on the plasma, but the $q=2$ region notably becomes an unstable red for $t_3$ relative to $t_1$, which may be interpreted as the region being at a location with a tearing risk.

A key question we seek to answer with Shapley analysis is the impact of ECH deposition changes at the $q=2$ surface on TM stability. Despite most actuation being equal to the reference shot in Fig.~\ref{fig:shap_199597}, Fig. \ref{fig:shap_199606} shows a control shot where a TM does not appear despite it being predicted by the model, likely due to the real-time modifications to ECH deposition which increased the current drive at the $q=2$ surface. In the control shot in Fig. \ref{fig:shap_199606}, the two timesteps chosen are $t_1$, where the tearing risk is high but the ECCD has not yet driven much current in the $q=2$ surface, and $t_2$, where the ECCD has provided more current drive aimed at the $q=2$ surface and consequently causing the TM probability to drop.

Several profile effects are expected from changing the ECH deposition to target the $q=2$ surface. Primarily, we expect changed electron temperature due to heating, as seen in Fig.~\ref{fig:shap_199606}b). Because our ECH setup also provides current drive, the $q=2$ region has a bump in current density $j$. Finally, ECH has a density pumpout effect, which may explain the lower density at the $t_2$ off-axis and edge regions, although other factors will impact this too.

The increase in $T_e$ between $t_1$ and $t_2$ is subtle, but it causes a small destabilizing effect, as was observed for the reference shot. However the increase in $J_{ECCD}$ at the $q=2$ surface, causing the bump on j at $t_2$ has a small stabilizing effect seen in Fig. \ref{fig:shap_199606}d), as is intended from driving ECCD at the $q=2$ surface where 2/1 TMs appear. Finally, the lower density in Fig. \ref{fig:shap_199606}c) has an overall stabilizing effect, particularly in the core and edge regions. 

While the two $q$ profiles in Fig. \ref{fig:shap_199606}e) are too similar to draw conclusions on their shape, the overall stability of the profile drops by 0.06 after ECCD stabilization of the shot. Since the $q$ profile largely describes the scenario, this could suggest the shot is evolving into a tearing unstable regime, but is kept stable by other profile changes induced by the ECH deposition change. 
 
Overall we observe the three key profile changes from ECH deposition modification, namely localized current drive, electron heating, and density pumpout, have an important effect on TM stability, suggesting that changes in ECH deposition played a role in avoiding the TM in this shot. This analysis shows the insight that Shapley analysis can have on the underlying physics learned by machine learning models for TMs. It also provides information on the triggering mechanisms and causes of a TM, such as a rise in temperature and pressure while rotation remains low.

\subsection{Shapley analysis II: which profile features affect TM stability?}\label{sec:shap2}

\begin{figure}
    \centering
    \includegraphics[width=86mm]{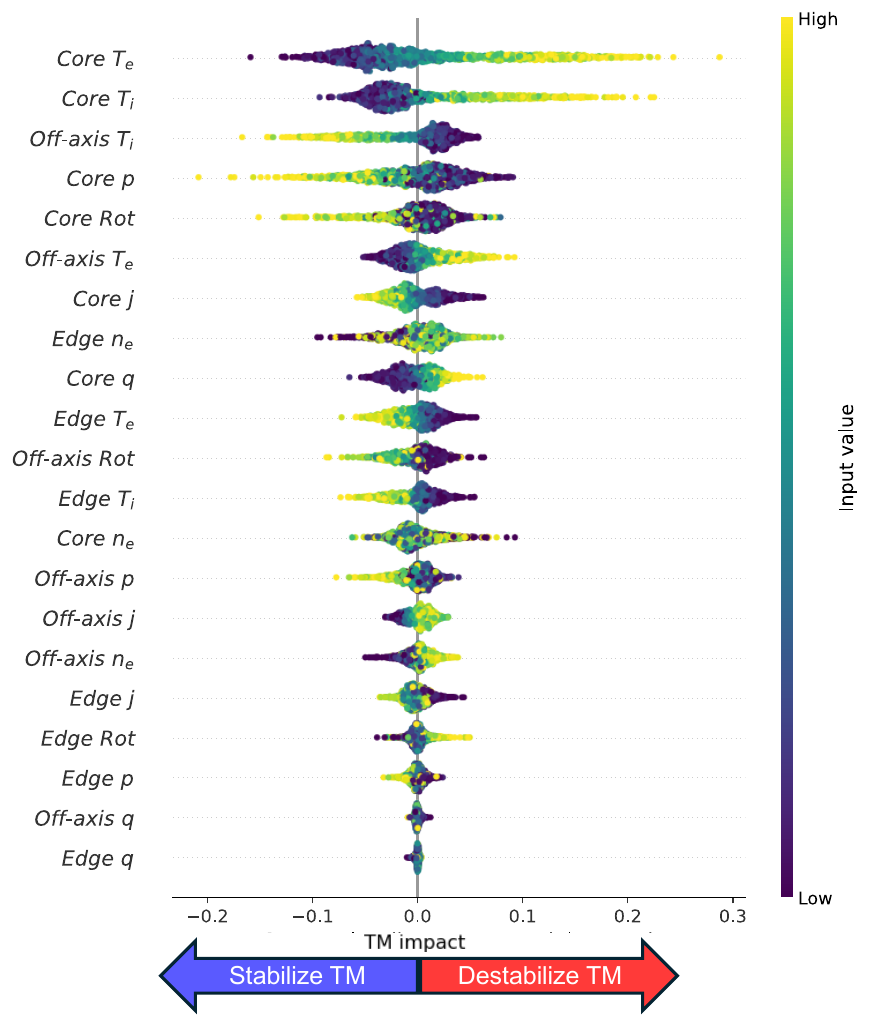}
    \caption{A general Shapley analysis of all inputs, ordered by influence on TM stability predictions in the dedicated experiment. Each row is a histogram of TM impact for a given value, with its color indicating the magnitude of the input. For example, core $T_e$ has the widest distribution suggesting high importance for TM prediction. As core $T_e$ increases (color becomes yellower), its TM impact clearly increases, showing that large core $T_e$ has a strong destabilizing effect.}
    \label{fig:beeswarm}
\end{figure}

Shapley analysis can be applied for a wider database analysis, providing insight on the overall impact of a profile feature, to draw more generalized conclusions. In this section we study which profile features have the largest impact in the scenario of our dedicated experiment. 

In Fig.~\ref{fig:beeswarm}, we plot the histograms of TM impact for each profile feature, with the 'core' region spanning $\psi_N\in[0,0.3]$, 'off-axis' region being $\psi_N\in[0.3,0.8]$ and 'edge' being $\psi_N\in[0.8,1]$ to represent the pedestal. The features are ordered by their overall TM impact, specifically by the mean of the absolute value of each TM impact, meaning the feature can be strongly stabilizing or destabilizing to TMs. This is visualized by the width of the histograms in the figure. By this metric, core $T_e$ is the highest and most important for TM stability prediction. It clearly shows that the higher the core $T_e$ value (or the lighter the color) the more destabilizing it will be to TMs. The same pattern is observed for core $T_i$, off-axis $T_e$ and core $q$ to a lesser extent. Notably, edge $T_e$ and $T_i$ display the opposite behavior, with a higher pedestal temperatures contributing to lower TM risk. 

The profiles largely derived from magnetic measurements, $j$ and $q$, have generally smaller TM impacts. This is likely due to all shots covered in this analysis having similar $j$ and $q$ profiles as they are in the same scenario. With less variation in $j$ and $q$ profiles we would expect smaller TM impact from these profiles. However, their core values show a clear pattern of high $j$ and therefore low $q$ leads to more TM stable shots, which is the expected result as stronger current drive should reduce TM risk. Real-time magnetic reconstructions used in this analysis also have lower accuracy, which may impact these results. 

\begin{figure}
    \centering
    \includegraphics[width=86mm]{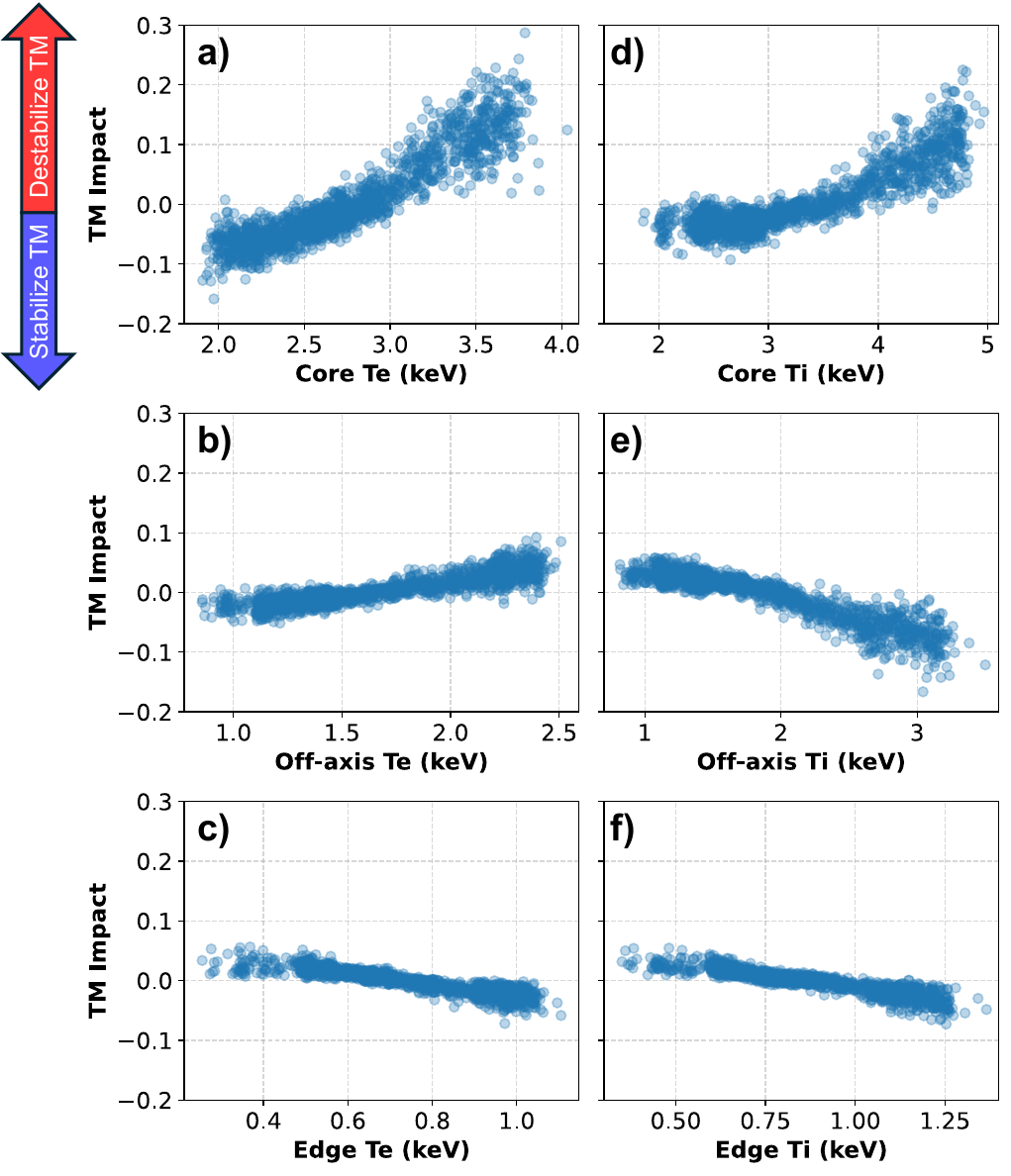}
    \caption{Scatter plots of TM impact from $T_e$ and $T_i$ across core, off-axis and edge regions, with each point representing one timeslice in the experimental data. From these plots, we can draw conclusions on the stability impact of different profile regions.}
    \label{fig:core-trends}
\end{figure}

The core and off-axis rotation are strongly stabilizing, as is expected from large flows suppressing MHD instabilities. However the core rotation distribution in Fig.~\ref{fig:beeswarm} has a dark center with light edges, suggesting that large core rotation can occasionally destabilize the plasma, and a low core rotation does not always strongly destabilize. Finally, we see a high edge rotation has an opposing, destabilizing effect, which indicates the importance of rotation gradients to stability.

While most profile features shown in Fig.~\ref{fig:beeswarm} show a clear pattern in color, it is important to highlight those with a larger scatter in color, particularly $n_e$ regions and some $p$ regions. This may be a result of low measurement accuracy, or strong correlations with other features affecting Shapley calculations. All the input profiles are of course highly correlated in a tokamak plasma, with the strongest correlation here being pressure, which is the product of temperature and density. Strongly correlated inputs should not significantly affect ML model performance, but does make the Shapley analysis more difficult, so it is important to remove such features before analysis in future applications. 

Focusing on the specific profile features of $T_e$ and $T_i$ in Fig.~\ref{fig:core-trends}, we see a clear difference in the pattern between the two profiles. While the TM risk due to core $T_e$ rises linearly with its magnitude, core $T_i$ shows little change in TM risk between 2keV and 3.5keV and an exponential rise at higher values. Both edge $T_e$ and $T_i$ show that a higher pedestal temperature is more stable to TMs. Finally the off-axis region displays opposite effects between $T_e$ and $T_i$, notably the off-axis $T_i$ in panel \textbf{e)} strongly stabilizing TMs as its magnitude increases. This shows an underlying difference in the mechanism linking $T_e$ and $T_i$ to TM onset in this scenario and suggests that broader, flatter temperature profiles are more beneficial for TM stability, especially in the $T_i$ profile. 

These insights can be used to inform future experiments to minimize TM risk without sacrificing performance. For example, the differences between $T_e$ and $T_i$ in TM risk suggest that the ratio between ECH (electron heating) and NBI (ion heating) fraction should be tuned to avoid the exponential rise in $T_i$-induced TM risk. A flatter $T_i$ profile is also found to stabilize TMs, so neutral beams aimed off-axis may result in better passive TM stability than typical on-axis NBI injection. 

\section{Conclusion}\label{sec:conclusion}

Using Shapley analysis, we were able to analyze the plasma profiles to understand their effects on TM stability predictions through a deep learning ML model. This was enabled by an accurate, long time horizon TM predictor model that was developed for DIII-D and proven in experiment to accurately predict TMs in real-time. This analysis validated generally understood physics observations such as higher $T_e$ destabilizing TMs while higher rotation stabilizes them. The analysis also led to new TM stability observations such as the TM risk increasing exponentially with core $T_i$, compared to a linear increase with core $T_e$, which suggests high $T_e/T_i$ fraction plasmas may be more stable to TMs. This analysis was performed on an elevated $q_{min}$ experiment and therefore only reflects the physics of that scenario, but a larger database study of different scenarios and machines may help uncover the key factors in developing tearing-free plasmas. 

Our Shapley analysis framework is also well-suited for analyzing various scenarios in DIII-D. By selecting an appropriate reference distribution, we can tailor the analysis to focus on specific plasma categories, such as advanced non-inductive plasmas, ITER baseline scenario plasmas, or other relevant scenarios. This filtering  simplifies the interpretation of the underlying physics because driving TM factors are known to be scenario-dependent.

Many experimental fusion phenomena are challenging to explain using physics-based models, making ML models an attractive alternative due to their high prediction accuracy. With the growing application of ML in fusion experiments, such as Alfvén eigenmodes, ELMs, and general disruptions, it has become more important than ever to understand how these models arrive at their predictions. By applying the Shapley analysis framework introduced in this paper, we can uncover the underlying physics and identify key features that should be controlled to prevent these instabilities, thereby improving the reliability and safety of fusion devices.

\section*{Acknowledgments}
This material is based upon work supported by the U.S. Department of Energy, Office of Science, Office of Fusion Energy Sciences, using the DIII-D National Fusion Facility, a DOE Office of Science user facility, under Award DE-FC02-04ER54698. Additionally, this material is supported by the U.S. Department of Energy, under Award DE-SC0015480. 

\section*{Disclaimer}

This report was prepared as an account of work sponsored by an agency of the United States Government. Neither the United States Government nor any agency thereof, nor any of their employees, makes any warranty, express or implied, or assumes any legal liability or responsibility for the accuracy, completeness, or usefulness of any information, apparatus, product, or process disclosed, or represents that its use would not infringe privately owned rights. Reference herein to any specific commercial product, process, or service by trade name, trademark, manufacturer, or otherwise does not necessarily constitute or imply its endorsement, recommendation, or favoring by the United States Government or any agency thereof. The views and opinions of authors expressed herein do not necessarily state or reflect those of the United States Government or any agency thereof.

\section*{References}

\bibliographystyle{unsrt}
\bibliography{TM-survival}

\begin{thebibliography}{10}

\bibitem{morosohkAcceleratedVersionNUBEAM2021}
Shira~M. Morosohk, Mark~D. Boyer, and Eugenio Schuster.
\newblock Accelerated version of {NUBEAM} capabilities in {DIII}-{D} using neural networks.
\newblock {\em Fusion Engineering and Design}, 163:112125, February 2021.

\bibitem{boyerRealtimeCapableModeling2019b}
M.D. Boyer, S.~Kaye, and K.~Erickson.
\newblock Real-time capable modeling of neutral beam injection on {NSTX}-{U} using neural networks.
\newblock {\em Nuclear Fusion}, 59(5):056008, May 2019.

\bibitem{morosohkNeuralNetworkModel2021}
S.M. Morosohk, A.~Pajares, T.~Rafiq, and E.~Schuster.
\newblock Neural network model of the multi-mode anomalous transport module for accelerated transport simulations.
\newblock {\em Nuclear Fusion}, 61(10):106040, October 2021.

\bibitem{rothsteinInitialTestingAlfven2024}
Andrew Rothstein, Azarakhsh Jalalvand, Joseph Abbate, Keith Erickson, and Egemen Kolemen.
\newblock Initial testing of {Alfvén} eigenmode feedback control with machine-learning observers on {DIII}-{D}.
\newblock {\em Nuclear Fusion}, 64(9):096020, September 2024.

\bibitem{keithRiskAwareFrameworkDevelopment2024a}
Zander Keith, Chirag Nagpal, Cristina Rea, and R.~Alex Tinguely.
\newblock Risk-{Aware} {Framework} {Development} for {Disruption} {Prediction}: {Alcator} {C}-{Mod} and {DIII}-{D} {Survival} {Analysis}.
\newblock {\em Journal of Fusion Energy}, 43(1):21, June 2024.

\bibitem{seoMultimodalPredictionTearing2023b}
Jaemin Seo, Rory Conlin, Andrew Rothstein, SangKyeun Kim, Joseph Abbate, Azarakhsh Jalalvand, and Egemen Kolemen.
\newblock Multimodal {Prediction} of {Tearing} {Instabilities} in a {Tokamak}.
\newblock In {\em 2023 {International} {Joint} {Conference} on {Neural} {Networks} ({IJCNN})}, pages 1--8, Gold Coast, Australia, June 2023. IEEE.

\bibitem{aiDisruptionPredictorBased}
X~K Ai, W~Zheng, M~Zhang, Y~H Ding, D~L Chen, Z~Y Chen, C~S Shen, B~H Guo, N~C Wang, Z~J Yang, Z~P Chen, Y~Pan, B~Shen, and B~J Xiao.
\newblock Disruption {Predictor} {Based} on {Convolutional} {Autoencoder} and {Its} {Cross}-tokamak {Deployment} {Strategy} {Study}.

\bibitem{reaRealtimeMachineLearningbased2019a}
C.~Rea, K.J. Montes, K.G. Erickson, R.S. Granetz, and R.A. Tinguely.
\newblock A real-time machine learning-based disruption predictor in {DIII}-{D}.
\newblock {\em Nuclear Fusion}, 59(9):096016, September 2019.

\bibitem{jalalvandAlfvenEigenmodeClassification2022}
Azarakhsh Jalalvand, Alan~A. Kaptanoglu, Alvin~V. Garcia, Andrew~O. Nelson, Joseph Abbate, Max~E. Austin, Geert Verdoolaege, Steven~L. Brunton, William~W. Heidbrink, and Egemen Kolemen.
\newblock Alfvén eigenmode classification based on {ECE} diagnostics at {DIII}-{D} using deep recurrent neural networks.
\newblock {\em Nuclear Fusion}, 62(2):026007, February 2022.

\bibitem{churchillDeepConvolutionalNeural2020a}
R.~M. Churchill, B.~Tobias, Y.~Zhu, and {DIII-D team}.
\newblock Deep convolutional neural networks for multi-scale time-series classification and application to tokamak disruption prediction using raw, high temporal resolution diagnostic data.
\newblock {\em Physics of Plasmas}, 27(6):062510, June 2020.

\bibitem{seoAvoidingFusionPlasma2024b}
Jaemin Seo, SangKyeun Kim, Azarakhsh Jalalvand, Rory Conlin, Andrew Rothstein, Joseph Abbate, Keith Erickson, Josiah Wai, Ricardo Shousha, and Egemen Kolemen.
\newblock Avoiding fusion plasma tearing instability with deep reinforcement learning.
\newblock {\em Nature}, 626(8000):746--751, February 2024.

\bibitem{degraveMagneticControlTokamak2022a}
Jonas Degrave, Federico Felici, Jonas Buchli, Michael Neunert, Brendan Tracey, Francesco Carpanese, Timo Ewalds, Roland Hafner, Abbas Abdolmaleki, Diego De~Las~Casas, Craig Donner, Leslie Fritz, Cristian Galperti, Andrea Huber, James Keeling, Maria Tsimpoukelli, Jackie Kay, Antoine Merle, Jean-Marc Moret, Seb Noury, Federico Pesamosca, David Pfau, Olivier Sauter, Cristian Sommariva, Stefano Coda, Basil Duval, Ambrogio Fasoli, Pushmeet Kohli, Koray Kavukcuoglu, Demis Hassabis, and Martin Riedmiller.
\newblock Magnetic control of tokamak plasmas through deep reinforcement learning.
\newblock {\em Nature}, 602(7897):414--419, February 2022.

\bibitem{liuInterpretableNeuralNetworks2023}
Zhuoyang Liu and Feng Xu.
\newblock Interpretable neural networks: principles and applications.
\newblock {\em Frontiers in Artificial Intelligence}, 6:974295, October 2023.

\bibitem{lundbergUnifiedApproachInterpreting2017}
Scott~M Lundberg and Su-In Lee.
\newblock A {Unified} {Approach} to {Interpreting} {Model} {Predictions}.
\newblock In I.~Guyon, U.~Von Luxburg, S.~Bengio, H.~Wallach, R.~Fergus, S.~Vishwanathan, and R.~Garnett, editors, {\em Advances in {Neural} {Information} {Processing} {Systems}}, volume~30. Curran Associates, Inc., 2017.

\bibitem{olofssonEventHazardFunction2018a}
K~E~J Olofsson, D~A Humphreys, and R~J~La Haye.
\newblock Event hazard function learning and survival analysis for tearing mode onset characterization.
\newblock {\em Plasma Physics and Controlled Fusion}, 60(8):084002, August 2018.

\bibitem{olofssonHazardFunctionExploration2019}
K.E.J. Olofsson, B.S. Sammuli, and D.A. Humphreys.
\newblock Hazard function exploration of tokamak tearing mode stability boundaries.
\newblock {\em Fusion Engineering and Design}, 146:1476--1479, September 2019.

\bibitem{fuMachineLearningControl2020c}
Yichen Fu, David Eldon, Keith Erickson, Kornee Kleijwegt, Leonard Lupin-Jimenez, Mark~D. Boyer, Nick Eidietis, Nathaniel Barbour, Olivier Izacard, and Egemen Kolemen.
\newblock Machine learning control for disruption and tearing mode avoidance.
\newblock {\em Physics of Plasmas}, 27(2):022501, February 2020.

\bibitem{Olofsson_Akçay_Sammuli_2022}
K.E.J. Olofsson, C.~Akçay, and B.S. Sammuli.
\newblock Database-wide hazard modelling of the onset of diii-d tearing modes with field features.
\newblock {\em Journal of Plasma Physics}, 88(5):895880503, 2022.

\bibitem{nagpalDeepSurvivalMachines2021c}
Chirag Nagpal, Xinyu Li, and Artur Dubrawski.
\newblock \textit{{Deep} {Survival} {Machines}} : {Fully} {Parametric} {Survival} {Regression} and {Representation} {Learning} for {Censored} {Data} {With} {Competing} {Risks}.
\newblock {\em IEEE Journal of Biomedical and Health Informatics}, 25(8):3163--3175, August 2021.

\bibitem{lundbergExplainableMachinelearningPredictions2018}
Scott~M. Lundberg, Bala Nair, Monica~S. Vavilala, Mayumi Horibe, Michael~J. Eisses, Trevor Adams, David~E. Liston, Daniel King-Wai Low, Shu-Fang Newman, Jerry Kim, and Su-In Lee.
\newblock Explainable machine-learning predictions for the prevention of hypoxaemia during surgery.
\newblock {\em Nature Biomedical Engineering}, 2(10):749--760, October 2018.

\bibitem{landremanHowDoesIon2025a}
Matt Landreman, Jong~Youl Choi, Caio Alves, Prasanna Balaprakash, R.~Michael Churchill, Rory Conlin, and Gareth Roberg-Clark.
\newblock How does ion temperature gradient turbulence depend on magnetic geometry? {Insights} from data and machine learning, February 2025.
\newblock arXiv:2502.11657 [physics].

\bibitem{pyragiusApplicationInterpretableMachine2024}
Tadas Pyragius, Cary Colgan, Hazel Lowe, Filip Janky, Matteo Fontana, Yichen Cai, and Graham Naylor.
\newblock Application of interpretable machine learning for cross-diagnostic inference on the {ST40} spherical tokamak, July 2024.
\newblock arXiv:2407.18741 [physics].

\bibitem{bardoczi_onset_2023}
L.~Bardóczi, N.J. Richner, and N.C. Logan.
\newblock The onset distribution of rotating m , n = 2 , 1 tearing modes and its consequences on the stability of high-confinement-mode plasmas in {DIII}-{D}.
\newblock {\em Nuclear Fusion}, 63(12):126052, December 2023.

\bibitem{bardoczi_empirical_2023}
L.~Bardóczi, N.~J. Richner, J.~Zhu, C.~Rea, and N.~C. Logan.
\newblock Empirical probability and machine learning analysis of \textit{m} , \textit{n} = 2, 1 tearing mode onset parameter dependence in {DIII}-{D} {H}-mode scenarios.
\newblock {\em Physics of Plasmas}, 30(9):092505, September 2023.

\bibitem{turcoCausesDisruptiveTearing2018}
F.~Turco, T.C. Luce, W.~Solomon, G.~Jackson, G.A. Navratil, and J.M. Hanson.
\newblock The causes of the disruptive tearing instabilities of the {ITER} {Baseline} {Scenario} in {DIII}-{D}.
\newblock {\em Nuclear Fusion}, 58(10):106043, October 2018.

\bibitem{bardocziRootCauseDisruptive2024}
L.~Bardoczi, N.J. Richner, N.C. Logan, E.J. Strait, C.T. Holcomb, J.~Zhu, and C.~Rea.
\newblock The root cause of disruptive {NTMs} and paths to stable operation in {DIII}-{D} {ITER} baseline scenario plasmas.
\newblock {\em Nuclear Fusion}, 64(12):126005, December 2024.

\bibitem{richnerUseDifferentialPlasma2024a}
N.J. Richner, L.~Bardóczi, J.D. Callen, R.J. La~Haye, N.C. Logan, and E.J. Strait.
\newblock Use of differential plasma rotation to prevent disruptive tearing mode onset from 3-wave coupling.
\newblock {\em Nuclear Fusion}, 64(10):106036, October 2024.

\bibitem{glasserIdealMHDDW2020a}
Alexander~S. Glasser, A.~H. Glasser, Rory Conlin, and Egemen Kolemen.
\newblock An ideal {MHD} \textit{$\delta${W}} stability analysis that bypasses the {Newcomb} equation.
\newblock {\em Physics of Plasmas}, 27(2):022114, February 2020.

\bibitem{carlstrom_design_1992}
T.~N. Carlstrom, G.~L. Campbell, J.~C. DeBoo, R.~Evanko, J.~Evans, C.~M. Greenfield, J.~Haskovec, C.~L. Hsieh, E.~McKee, R.~T. Snider, R.~Stockdale, P.~K. Trost, and M.~P. Thomas.
\newblock Design and operation of the multipulse {Thomson} scattering diagnostic on {DIII}-{D} (invited).
\newblock {\em Review of Scientific Instruments}, 63(10):4901--4906, October 1992.

\bibitem{seraydarian_multichordal_1986}
Raymond~P. Seraydarian and Keith~H. Burrell.
\newblock Multichordal charge-exchange recombination spectroscopy on the {DIII}-{D} tokamak.
\newblock {\em Review of Scientific Instruments}, 57(8):2012--2014, August 1986.

\bibitem{wroblewski_motional_1990}
D.~Wróblewski, K.~H. Burrell, L.~L. Lao, P.~Politzer, and W.~P. West.
\newblock Motional {Stark} effect polarimetry for a current profile diagnostic in {DIII}-{D}.
\newblock {\em Review of Scientific Instruments}, 61(11):3552--3556, November 1990.

\bibitem{lao_reconstruction_1985}
L.L. Lao, H.~St.~John, R.D. Stambaugh, A.G. Kellman, and W.~Pfeiffer.
\newblock Reconstruction of current profile parameters and plasma shapes in tokamaks.
\newblock {\em Nuclear Fusion}, 25(11):1611--1622, November 1985.

\bibitem{sweeney_relationship_2018}
R.~Sweeney, W.~Choi, M.~Austin, M.~Brookman, V.~Izzo, M.~Knolker, R.J. La~Haye, A.~Leonard, E.~Strait, F.A. Volpe, and {The DIII-D Team}.
\newblock Relationship between locked modes and thermal quenches in {DIII}-{D}.
\newblock {\em Nuclear Fusion}, 58(5):056022, May 2018.

\bibitem{shoushaMachineLearningbasedRealtime2023}
Ricardo Shousha, Jaemin Seo, Keith Erickson, Zichuan Xing, SangKyeun Kim, Joseph Abbate, and Egemen Kolemen.
\newblock Machine learning-based real-time kinetic profile reconstruction in {DIII}-{D}.
\newblock {\em Nuclear Fusion}, 64(2):026006, 2023.
\newblock Publisher: IOP Publishing.

\bibitem{xingCAKEConsistentAutomatic2021}
Z.A. Xing, D.~Eldon, A.O. Nelson, M.A. Roelofs, W.J. Eggert, O.~Izacard, A.S. Glasser, N.C. Logan, O.~Meneghini, S.P. Smith, R.~Nazikian, and E.~Kolemen.
\newblock {CAKE}: {Consistent} {Automatic} {Kinetic} {Equilibrium} reconstruction.
\newblock {\em Fusion Engineering and Design}, 163:112163, February 2021.

\bibitem{victorGLOBALSTABILITYELEVATEDQMINb}
B~S Victor, C~T Holcomb, K~E Thome, W~P Wehner, C~S Collins, C~C Petty, San Diego, J~M Park, J~M Hanson, B~A Grierson, M~W Aslin, and South Hadley.
\newblock {GLOBAL} {STABILITY} {OF} {ELEVATED}-{QMIN}, {STEADY}-{STATE} {SCENARIO} {PLASMAS} {ON} {DIII}-{D}.

\bibitem{rothsteinaHighqmin}
A.~Rothstein, H.~Farre-Kaga, S.K. Kim, K.~Erickson, and E.~Kolemen.
\newblock Preemptive tearing mode stabilization with multi-tasking ech in advanced tokamak plasmas.
\newblock In process for publication.

\bibitem{kolemenStateoftheartNeoclassicalTearing2014d}
E.~Kolemen, A.S. Welander, R.J. La~Haye, N.W. Eidietis, D.A. Humphreys, J.~Lohr, V.~Noraky, B.G. Penaflor, R.~Prater, and F.~Turco.
\newblock State-of-the-art neoclassical tearing mode control in {DIII}-{D} using real-time steerable electron cyclotron current drive launchers.
\newblock {\em Nuclear Fusion}, 54(7):073020, July 2014.

\end{thebibliography}

\newpage
\appendix
\section{Detailed considerations on database processing and TM labelling}
\label{app:data_processing}
\begin{itemize}
    \item $I_P$ rampup and rampdown data is excluded, as we wished to predict flattop TMs which can be controlled by changing ECH deposition. We wouldn't want to change ECH deposition during rampup as it may affect the scenario.
    \item A TM was considered to have occurred if the amplitude of the root mean square (RMS) of the n=1 magnetic fluctuations, called n1rms, signal peaked above 12G for a continuous 50ms along with additional constraints on $H_{98}$ and $q_{95}$ to only consider H-modes plasmas. The onset time of the TM was determined to be when the n1rms first reached 10\% of the peak n1rms signal. This means higher m number modes were also included, although they are less likely than the intended 2/1 modes. It also means that n=2,3 modes are ignored.
    \item The n1,2,3rms signals, the RMS magnetic signal for n=1,2,3 modes, are excluded from the training set to avoid the model overly relying on these signals, as they are used for labeling.
    \item The data was chosen to be every 20ms as this is enough time for diagnostics and EFITs to yield updated results, faster than $\tau_R$ and $\tau_E$ ensuring the profile are equilibrated, but not too fast that the model overfits to noise. Actuation such as modifying ECH deposition and NBI power adjustment will also affect the plasma on the order of $100ms$, so this is a good compromise.
    \item The NBI power and torque signals were smoothed to remove the modulation spikes which are too fast to affect the plasma equilibrium. 
    \item ECH mirror angles are not included as an input. The model is designed to observe physics quantities such as $q$ profile changes rather than actuation quantities. This would ensure the models weren’t biased to predicting ECCD suppressed TMs, but it should be a learned effect on the profiles. DIII-D also has limited variety in ECH mirror angles, making it difficult for models to learn relations between stability and mirror angle. 
    \item The model did not have memory, which means all the diagnostic noise would appear in each prediction timestep, affecting TM predictions. We therefore apply a low pass filter to the outputs to ensure noise spikes don’t cause false positives.
    \item All the inputs are signals available in real-time in DIII-D, and could be available in most tokamaks. This is to ensure that such a model is applicable to real-time control in present and future reactors. 
    \item The shot time is not provided to the model to avoid overfitting to specific times. Certain scenarios frequently have TMs at specific times, which may cause the model to overfit to time and not learn from the profiles. 
\end{itemize}

\section{Event characterization details}\label{app:event-char}
\begin{figure}
    \centering
    \includegraphics[width=86mm]{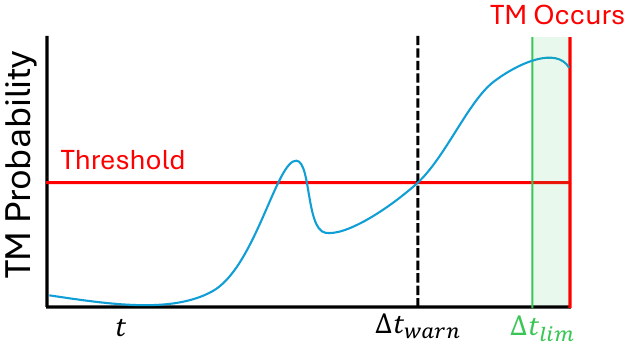}
    \caption{Diagram of an example TM prediction. $\Delta t_{warn}$ is the warning time for a TM and $\Delta t_{lim}$ the time considered too late to be considered a correct prediction.}
    \label{fig:event_label}
\end{figure}

Fig.~\ref{fig:event_label} depicts an example TM prediction result to explain the event labelling definitions used for our database. This is a true positive because the threshold is crossed and a TM occurs, however we define the $\Delta t_{warn}$ as the \textit{last} time the threshold is crossed. Considering the \textit{first} time it is crossed results in inflated warning time statistics. 

If the last time the threshold is crossed comes after $\Delta t_{lim}$, this is considered a false negative even if the event is correctly predicted, since it is too late to actuate on the TM. This paper uses $\Delta t_{lim} = 100ms$ as this is approximates the time needed for ECH and most actuation to affect the plasma. Only 21/1476 shots with TMs were in this category because of our high warning times.

\section{Shapley Toy Model}\label{app:toy-model}
We consider a toy model to determine the key factors affecting a football team's win percentage: 

\begin{align*}
\textrm{winRate} = \frac{1}{\textrm{norm}}\left[ 2\cdot (\textrm{Cost}_\textrm{squad})^2  - 20\cdot (\textrm{Num}_\textrm{injury}) \right . &\\
    \left. - 10\cdot (\textrm{Age}_\textrm{avg}- 24)^2\right]&
\end{align*}
With this exact formula, we can exactly see how each term, average cost $\textrm{Cost}_\textrm{squad}$, age $\textrm{Age}_\textrm{avg}$, and injury count $\textrm{Num}_\textrm{injury}$, contribute to the calculated win rate. However for our toy model Shapley analysis, this function is hidden and we consider it as a black-box model that takes input $(\textrm{Cost}_\textrm{squad},\textrm{Age}_\textrm{avg},\textrm{Num}_\textrm{injury})$ and outputs the team's win rate.

\begin{table}
    \begin{minipage}[t]{0.4\textwidth}
        \raggedright
        \fontsize{7pt}{11pt}\selectfont
        \begin{tabular}{lcc}
            \toprule
            \textbf{Team} & \textbf{Team 1} & \textbf{Team 2} \\
            \hline 
            Cost (millions)   & \$25       & \$24       \\
            Age (years)    & 22    & 23    \\
            Injury Count  & 8       & 5       \\
            Win rate      & 0.65     & 0.62   \\
            \bottomrule
        \end{tabular}
        \label{tab:team_comparison}
    \end{minipage}
    \hspace{0.7cm}
    \begin{minipage}[t]{0.4\textwidth}
        \raggedleft
        \fontsize{7pt}{11pt}\selectfont
        \begin{tabular}{lcc}
            \toprule
            \textbf{Context} & \textbf{Professional} & \textbf{Amateur} \\
            \hline 
            Cost (millions)   & \$15 - \$30         & \$13 - \$15         \\
            Age (years)    & 20 - 30     & 20 - 30     \\
            Injury Count  & 0 - 10         & 0 - 10         \\
            \bottomrule
        \end{tabular}
        \label{tab:level_comparison}
    \end{minipage}
    \caption{Left table: Example Teams 1 and 2  average cost, age, injury counts and the corresponding win rate. Right table: professional and amateur background distributions used to calculate Shapley values, where the only difference is the range of squad costs as professionals cost much more than amateurs. }
\end{table}

\begin{figure}
    \centering
    \includegraphics[width=86mm]{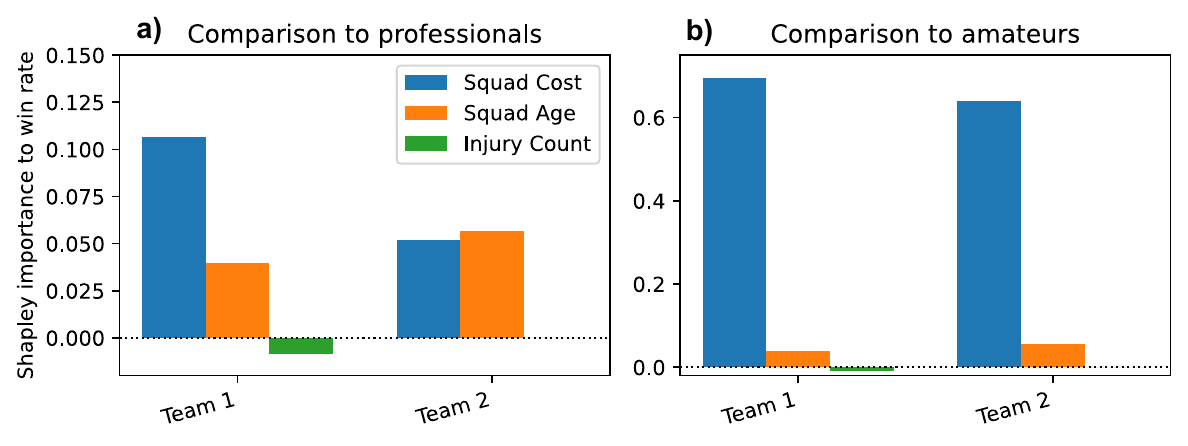}
    \caption{Shapley values for our toy model. \textbf{a)} Shapley values between two professional teams show meaningful contributions from Cost, Age, and Injury Count, where the difference in Cost is the driving factor. \textbf{b)} Comparing to a background of Amateur teams, the cost is now the sole driving factor for the team's win rates and age and injury counts are negligible. }
    \label{fig:shap_football}
\end{figure}

Fig.~\ref{fig:shap_football} shows the Shapley analysis on the win rate model using two background distributions (professional with higher costs vs amateur) and two professional teams. In a), the higher squad cost in Team 1 gives it a larger Shapley value. The higher value of squad cost ($25M$) leads to a 0.1 improvement in the win rate prediction relative to other professionals, which intuitively makes sense as the cost of the squad has quadratic dependence on the win rate. 

Next we see the importance of the reference distribution when comparing Fig. \ref{fig:shap_football} a) to b). When compared to other professionals, amateur age and injuries will have a meaningful effect on a team's win rate. However, when compared to amateurs that have a significantly lower cost, the professional team's superior cost dominates all other factors. Note that the total Shapley value in b) is larger than in a) because Shapley values are relative to the distribution average, which is significantly lower in the amateur group. 
\end{document}